\documentclass[journal,twocolumn]{IEEEtran}
\usepackage{cite}
\usepackage[pdftex]{graphicx}
\usepackage{amsmath}
\usepackage{algorithmic}
\usepackage{xcolor}
\usepackage{array}
\usepackage{algorithm,algorithmic}
\usepackage[utf8]{inputenc}
\usepackage{enumitem} 
\usepackage[many]{tcolorbox}
\usepackage{epsfig} 
\usepackage{pdfpages} 
\usepackage{amsmath, amssymb, mathrsfs}
\usepackage{epstopdf}

\usepackage{color}
\usepackage{subfig}

\usepackage{cite}

\usepackage{filecontents}
\usepackage{graphicx}
\usepackage{float}
\usepackage{subfig}

\newcommand\mydots{\hbox to 1em{.\hss.\hss.}}

\begin{document}

\title{Stress-informed Control of Medium- and High-head Hydropower Plants to Reduce Penstock Fatigue}

%
%

\author{Stefano~Cassano
        and~Fabrizio~Sossan
\thanks{S. Cassano and F. Sossan are with the Centre for processes, renewable energies and energy systems (PERSEE) of MINES ParisTech, Sophia Antipolis, France. E-mail: {stefano.cassano, fabrizio.sossan}@mines-paristech.fr.}
\thanks{Research supported by the European Union Horizon 2020 research and innovation program under XFLEX HYDRO, grant agreement 857832.
}
}



\maketitle

\begin{abstract}
The displacement of conventional generation in favor of stochastic renewable requires increasing regulation duties from the remaining dispatchable resources. In high- and medium-head hydropower plants (HPPs), providing regulation services to the grid and frequently changing the plant's set-point causes water hammer, which engenders pressure and stress transients within the pressurized conduits, especially the penstock, damaging it in the long run. This paper proposes a model predictive control (MPC) that explicitly models the hydraulic transients within the penstock. It achieves to reduce the mechanical loads on the penstock wall, and, consequently, fatigue effectively. Thanks to a suitable linearization of the plant model, the optimization problem underlying the MPC scheme is convex and can be solved with off-the-shelf optimization libraries.  The performance of the proposed controller is tested with numerical simulations on a 230~MW medium-head HPP with Francis turbine providing primary frequency control. Simulation results show substantially reduced penstock fatigue, existing approaches outperformed, and problem resolution times compatible with real-time control requirements.
\end{abstract}

\begin{IEEEkeywords}
Model predictive control, Hydropower plants, Frequency regulation.
\end{IEEEkeywords}


%
\IEEEpeerreviewmaketitle

\section{Introduction}
Hydropower is the leading renewable energy globally, supplying 70\% of all renewable electricity and  17\%  of  the  total electricity generation \cite{iea}. Besides generating electricity, hydropower plants (HPPs) provide fundamental regulation services to the power grid, ensuring its correct and reliable operations. However, the displacement of conventional dispatchable generation in favor of production from stochastic renewable sources (e.g., PV and wind) causes increasing regulation duties for the remaining dispatchable generation resources, which result in increased wear (due to more intense use) and tear (damage from mechanical fatigue) of the components\cite{yang_wear_2016}. 

In HPPs, wear-and-tear phenomena depend on the type of the plant and on its head, that is the difference in elevation between the upstream and the turbine. 

In medium- and high-head plants, the focus of this paper, the concern is the hydraulic pressurized conduit that feeds water to the turbine, the so-called \emph{penstock}. Indeed, sudden variations of the guide vane due to changing the plant's power output result in abrupt changes of the water pressure that reflect back and forth in the penstock (\emph{water hammer} effect). This phenomenon results in increased mechanical stress on the penstock wall, damaging it in the long run due to fatigue. 
Water hammer happens when the elastic behavior of the water conduit is not negligible anymore, and this happens with long penstock (and so with ”sufficiently” large heads). In medium-head hydropower plants (generally where the head is larger than 30-40 meters), penstocks are typically long enough to feature substantial water hammers.
In high-head hydropower plants (head larger than 200-300 meters), penstocks are longer and, therefore, one or more surge tanks may be included to reduce the over-pressures in the piping system.
On the other hand, low-head (head
smaller than 30-40 meters) plants have short penstock and water hammer is typically not an issue, so is not penstock fatigue.
The works in \cite{ZHANG2019690, LUO2010192} have shown that, in such plants, wear and tear concerns are associated with the actuating mechanisms of the guide vane (which regulates the amount of water flowing to the turbine) and Kaplan turbine's blades (which adjusts the efficiency).

The existing literature related to penstock fatigue can be classified into two categories: structural design, and analysis of the penstock fatigue arising during plant operations.

In the first category, the authors of \cite{surgetank,en12132527} investigate the optimal design, location and response of surge tanks to reduce the peak pressure in the pipe system.
The work in \cite{inproceedings} shows how the penstock's wall thickness, internal section area and type of material can prevent destructive effects of the water hammer effect by reducing the pressure wave speed.

In the second category, the problem of penstock fatigue arising from increased power regulation duties was first acknowledged by Nicolet et al. in \cite{Nicolet2010EvaluationOP} that investigated the solicitations on the penstock resulting from the provision of secondary frequency regulation.
Dreyer et al. in \cite{Dreyer2019DigitalCF} estimated that providing primary and secondary frequency regulation can increase penstock fatigue by a factor 10, with obvious repercussion on the service life of this critical component.

Accrued penstock fatigue resulting from operations makes it of crucial interest to revisit existing hydropower plant controllers. Indeed, typical HPP controllers, known as governors, are configured to ensure that the nominal limits of the plants are respected, with no focus, however, on alleviating the long-term impact of increased regulation duties on fatigue. Wear-and-tear concerns for HPPs might be such that plant operators make conservative decisions about the regulation to provide, to the detriment, however, of grid support and profit they could make from electricity regulation markets.

Few works in the literature have addressed the topic of control methods to limit wear and tear in HPPs during frequency regulation. The work in \cite{9160666} proposed to low-pass filter the HPP power output set-point to avoid frequent changes of the guide vane while providing the remaining power with a battery energy storage system (BESS).
The authors of \cite{7514942} compare the widely adopted dead-zone filter for the grid frequency against a floating dead-band and a low-pass filter, pointing out the need to trade-off between the turbine's wear reduction and regulation performance.
Although low-pass and dead-band filters effectively reduce fatigue, they require manual tuning. Besides, and most importantly, they rely on the implicit and empirical assumption that decreasing guide vane movements reduces fatigue without being informed, however, of the actual mechanical loads on the penstock.

This paper proposes a model predictive control (MPC) to compute the power set-points of a medium-head HPP while explicitly accounting for the mechanical loads and engendered stress in the penstock. We model mechanical loads and fatigue of the penstock with linearized guide vane-to-load models from the literature \cite{ISGT}. By virtue of this feature, the problem is convex and can be solved with real-time requirements.

To the best of these authors' knowledge, this paper is the first attempt in the literature to formulate stress constraints explicitly in the plant control problem. The formulation of the MPC problem stands as the distinguishing contribution of this work.

The rest of this paper is organized as follows. Section II describes the problem and introduces the modeling requirements and methodology; Section III describes the formulation of MPC for fatigue reduction; Section IV and V describe the case study, benchmark methods, and the results; Section VI concludes the paper.

\begin{figure*}
\begin{center}
    \includegraphics[]{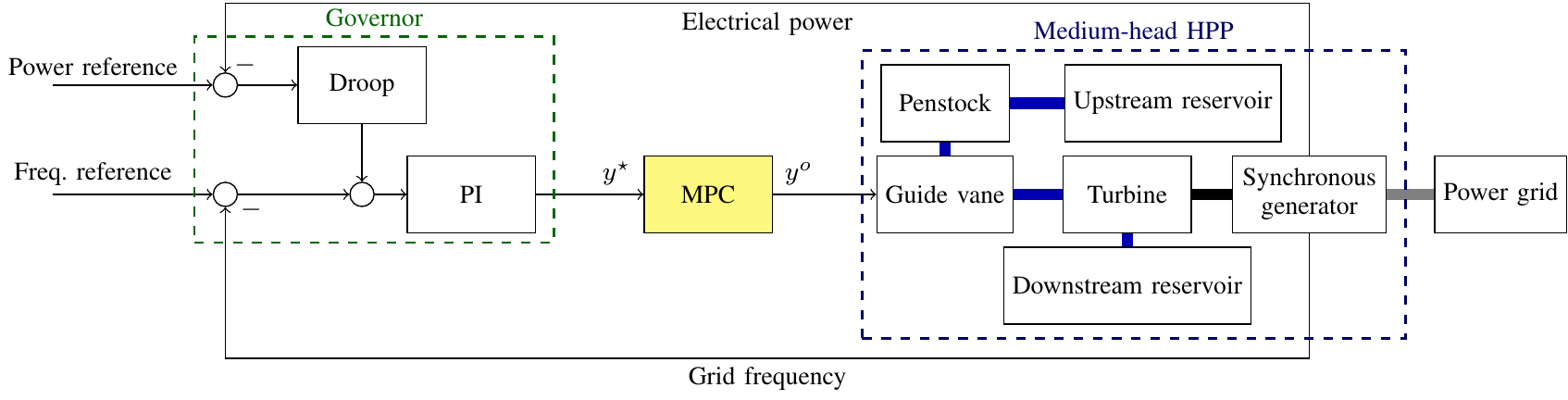}\\
\end{center}
\caption{Governor and main components of a medium-head HPP.
}\label{fig:hpp_diag}
\end{figure*}


\section{Problem statement and Modelling of HPPs}

\subsection{Problem statement}
Before illustrating the problem, the typical configuration of medium-head HPP is discussed with the objective of illustrating the application. The main components of a medium-head HPP (shown in the blue dashed rectangle of Fig.~\ref{fig:hpp_diag}) are the reservoirs, the penstock, the guide vane, the turbine, and the synchronous electrical generator. The guide vane regulates the water flow to the turbine. The electrical generator, connected to and synchronized with the power grid, converts mechanical power into electricity and supplies the grid. The plant governor (in the green dashed box of Fig.~\ref{fig:hpp_diag}) regulates the plant's power output by acting on the plant guide vane, whose set-point is denoted by $y^\star$.\footnote{High-head power plants rely on a similar configuration, with the addition of surge tanks, which can be modeled with the same principles discussed here.} The power reference signal in Fig.~\ref{fig:hpp_diag} is according to a production schedule (e.g., from electricity market commitments) or for secondary frequency regulation. The regulation loop on the grid frequency is a standard droop regulator for PFR. Governor parameters (i.e., droop and PI controller) are chosen to respect the static limits of the plant and timing of the plant power output. Set-point $y^\star$ does not account, however, for the accumulated effects of mechanical load variations in the penstock, and repercussions on the fatigue. 

The objective of this paper is to design a controller to reduce the mechanical fatigue in the penstock by properly "tapping" the guide vane set-point. More specifically, this controller (the yellow box in Fig.~\ref{fig:hpp_diag}) determines a new guide vane opening, $y^o$, with the following two properties:
\begin{enumerate}
    \item it should not result in mechanical loads engendering fatigue;
    \item the new guide vane reference should be as close as possible to $y^\star$. This requirement is to preserve the original regulation duties of the plant and avoid excess curtailment of the regulation duties, which could result in economic penalties or disqualification from regulation services.
\end{enumerate}

The controller is a receding horizon MPC, described in the next section. The mechanical loads in the penstock and the resulting stress are formulated with dynamic models that capture the water pressure dynamics within the conduits. Modeling water's pressure dynamics requires modeling all the hydraulic circuit components, most notably the penstock and the turbine. Even if not a contribution of this paper, their models are briefly described in the rest of this section with the objective of illustrating the required linearization in view of the formulation of the MPC problem.

\subsection{Fundamentals of hydropower plant modeling} \label{sec:Modelling}
HPP models for power systems studies are typically transfer function and equivalent circuit models. They are also referred to as one-dimensional models, as opposed to computational fluid dynamic (CFD) models in two or three dimensions, which are used to simulate the detailed behavior of a single hydraulic component. CFD models are generally not suited to control and power systems simulation applications due to their computational complexity. The equivalent circuit analogy adopted in this paper consists in modeling the piezometric head (or pressure), $H$, and water discharge, $Q$, at a given point of an hydraulic circuit as a voltage and current of an electrical circuit \cite{electric}.

The diagram of the equivalent circuit model of a medium-head HPP is shown in Fig.~\ref{fig:EEC}. It models the flow of water within the hydraulic circuit of the plant. The two voltage sources at the far ends of the circuit, $H_u$ and $H_d$, are the water level at the upstream and downstream reservoir, respectively. Then, the series of RLC circuits model the penstock, and the controlled voltage source $H_t$ models the turbine, as described next.

\subsubsection{Penstock model}
\begin{figure}
    \centering
    \includegraphics[width=1\columnwidth]{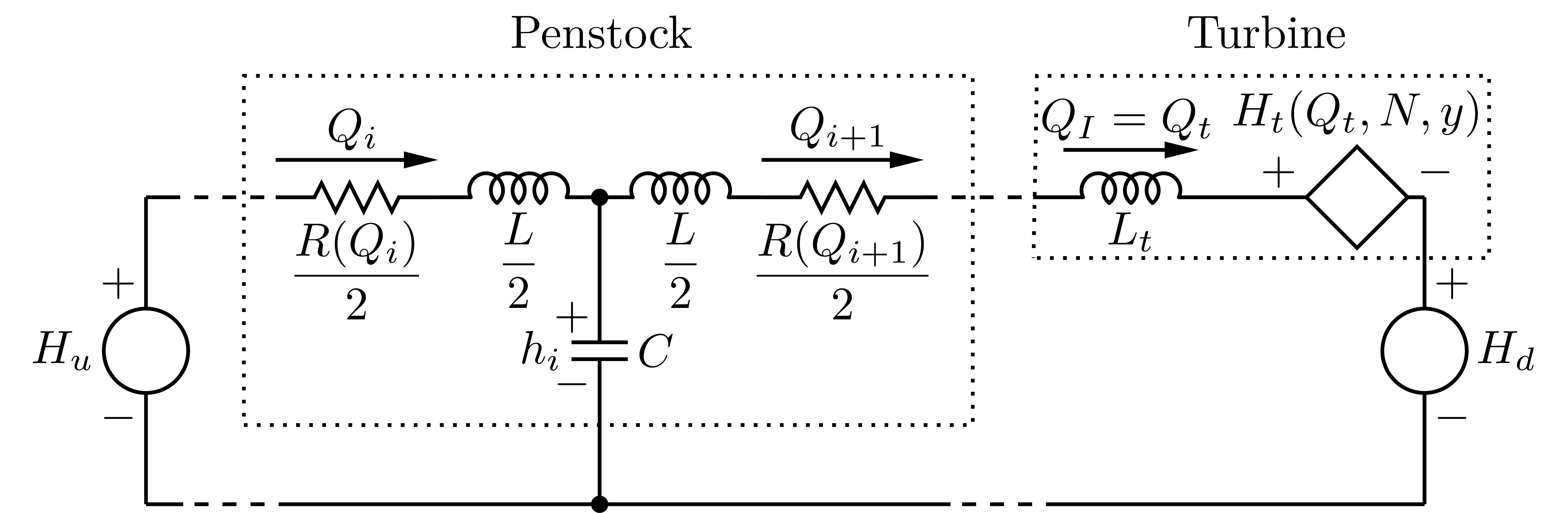}\\
    \caption{Equivalent circuit model of a medium-head HPP.}\label{fig:EEC}
\end{figure}

The penstock is modeled by discretizing the conduit in a finite number of elements, say $I$, where each element is modelled as a third-order RLC circuit, as shown Fig.~\ref{fig:EEC} with the index $i=1,\dots, I$. The voltage $h_i$ is the water head (or static pressure) in the central part of the penstock element, whereas $Q_i$ and $Q_{i+1}$ is the water flow in the receiving and sending end of it. This model allows to capture head losses along the conduit, and most importantly, pressure dynamics within the penstock, which are at the origin of the water hammer effect and induce mechanical fatigue on the penstock wall, as elaborated in Section~\ref{sec:fatigue}. 

The circuit parameters can be determined based on penstock's physical properties and are \cite{Nicolet:98534}:
\begin{align}
    R = \frac{\lambda\cdot |Q|\cdot dx}{2\cdot g\cdot D\cdot A^{2}}, && L = \frac{dx}{g\cdot A}, && C = \frac{g\cdot A\cdot dx}{a^{2}},
    \label{eqn:RLC}
\end{align}
where $dx$ is the spatial discretization interval, $\lambda$ is the Darcy-Weisbach friction coefficient, $g$ is the gravitational acceleration, $A$ and $D$ are the penstock's cross-section and diameter, respectively, and $a$ the wave speed in meters per second.

\subsubsection{Turbine}
An hydraulic turbine converts hydraulic energy into mechanical work. It is modeled as a variable voltage source, denoted by generator $H_{t}(\cdot)$ in Fig.~\ref{fig:EEC}. The voltage is derived by solving the turbine's characteristic curve, which determines the turbine head by relating all turbine's hydraulic and mechanical quantities, namely its rotational speed $N$, head $H_t$, and flow $Q_t$ and guide vane opening $y$. Turbine's characteristic curves are typically derived experimentally from the plant, see e.g. \cite{Nicolet:98534}, and are non-linear. This model assumes that the transient behavior of the turbine can be simulated as a succession of different steady-state operating points (“quasi-static” model). It ensures sufficient accuracy for all the flow regimes during transient and tractable computational times \cite{Nicolet:98534}. Finally, inertia effects of the water in the turbine are modeled through the equivalent inductance of the turbine $L_t$.


\subsection{Non-linear and linear state-space model}
The equivalent circuit model in Fig.~\ref{fig:EEC} can be synthesized as a state-space model, which is convenient for the MPC formulation. 

In order to achieve a tractable formulation of the MPC problem, we resort to the linearized model described in \cite{ISGT}, which are briefly summarized in the following for the sake of clarity. As shown in \cite{ISGT}, these linear models are capable of estimating the penstock head quite accurately. Thanks to their good estimation performance, they are suitable to implement stress constraints in the MPC problem with a satisfying accuracy level. The accuracy of these models for the case study of this paper is shown in the Results section.

A first approximation of the model in the prospect of its linearization is assuming a constant rotor and turbine pulsation at $\omega_0=2\pi\cdot f_0/p$, where $f_0$ is the nominal grid frequency and $p$ the polar couples of the electric generator, within the time interval. This assumption is deemed reasonable because grid frequency deviations during normal power system operations are small, being the grid frequency controlled and contained in a narrow band. This assumption will be verified in the simulations, which are performed considering a more accurate model than the one used in the MPC, including a generator and swing equation models. Because of this assumption, the turbine's rotational speed does not appear in the state vector, but it enters the model as a parameter.


The state vector of the model in Fig.~\ref{fig:EEC} is (bold typeface denotes vectors):
\begin{align}
\boldsymbol{x} &=\begin{bmatrix} Q_1 & \dots & Q_{I} & h_1 & \dots & h_I & Q_t \end{bmatrix}^\top \label{eq:state}
\end{align}
where $Q_i$ and $h_i$ for $i=1, \dots, I$ are, respectively, the water flow and head in the penstock element $i$, $Q_t$ is the turbine discharge and $^\top$ denotes transpose. Vector \eqref{eq:state} has dimension $2I+1$, with $I+1$ discharges, and $I$ heads.


The state-space model of the circuit in Fig.~\ref{fig:EEC} is non-linear because, i), the turbine characteristic curve is non-linear \cite{Nicolet:98534} and, ii), the hydroacoustic resistance $R$ in \eqref{eqn:RLC} depends on the flow, resulting in a bi-linear relationship between components of the state vector.

The linearization consists in i) assuming that the hydroacoustic resistance is independent of the flow, and ii) modeling the turbine with a first-order Taylor expansion around the operating point.

Under these assumptions, the model can be written as the following continuous time linear state-space (the notation $\tilde{\cdot}$ refers to continuous-time dynamics):
\begin{align}
& \dot{\boldsymbol{x}}\left(\tilde{t}\right) = \tilde{A} \boldsymbol{x}(\tilde{t}) + \tilde{B}_{y} y(\tilde{t}) + \tilde{B}_z  \begin{bmatrix} H_u(\tilde{t}) \\ \mu - H_d(\tilde{t}) \end{bmatrix} \label{eq:ss}
\end{align}
or, more compactly as
\begin{align}
& \dot{\boldsymbol{x}}(\tilde{t}) = \tilde{A} \boldsymbol{x}(\tilde{t}) + \tilde{B}_{y} y(\tilde{t}) + B_z  \boldsymbol{z}(\tilde{t})
\end{align}
where $\tilde{A} \in \mathbb{R}^{(2I + 1) \times (2I + 1)}$ is the system matrix, $\boldsymbol{x}$ is the time varying state vector in \eqref{eq:state}, $\tilde{B}_y \in \mathbb{R}^{(2I + 1) \times 1}$ the input matrix for the controllable input (i.e., guide vane $y$), $\tilde{B}_z \in \mathbb{R}^{(2I + 1) \times 2}$ the input matrix for the uncontrollable inputs in $\boldsymbol{z}(t)$, (i.e., the head of the up- and down-stream reservoir) and $\mu$ is an input coefficient that follows from the linearization procedure.

Quantities $\tilde{A}, \tilde{B}_y, \tilde{B}_z, \mu$ are parameters calculated based on the plant characteristics and linearization point, as described in \cite{ISGT}. They are an input of this problem. Up- and down-stream head $H_u, H_d$ are also input information from measurements. Since we target PFR (i.e., time dynamics in the order of seconds), reservoirs' water levels are assumed constant.


As discussed next, the fatigue constraints of the penstock are formulated as a function of the linearized head model, enabling a convex formulation of the optimization problem underlying MPC.

\subsection{Evaluating penstock fatigue}\label{sec:fatigue}
 Fatigue is a failure mechanism that occurs to materials due to accumulated mechanical loads (or stress). Fatigue failures are typically sudden and hard to predict due to the variability of the loads, point of application, direction, etc. In material science, this phenomenon is also known as "crack growth" and it deals with small imperfections in the material that grow due to loading over time, ultimately leading to ruptures. However, there exist model to estimate its evolution over time.
 
Fatigue and residual lifetime of a component can be modelled considering the numbers of cycles to failure for a specific level of loading, typically determined empirically on the basis of physical testing performed on material specimen, e.g., \cite{fatigue}. In this paper, we assess fatigue using the stress-life method based on Wohler's curve, discussed later in this section. This method is used in applications where the applied stress is within the elastic range of the material, and the material has a long cycle life (i.e., more than $10^4$ cycles to failure), as for the penstock. The stress-life approach is widely adopted for fatigue evaluation due to its simplicity. It is known to give conservative lifetime estimates compared to other methods, such as crack propagation theory \cite{MCBAGONLURI2005291, adam}, which will be considered in future works. However, as demonstrated in this paper, it provides an actionable way to determine operational patterns possibly conducive to excess fatigue.

\begin{figure}
    \centering
    \footnotesize
    \includegraphics[]{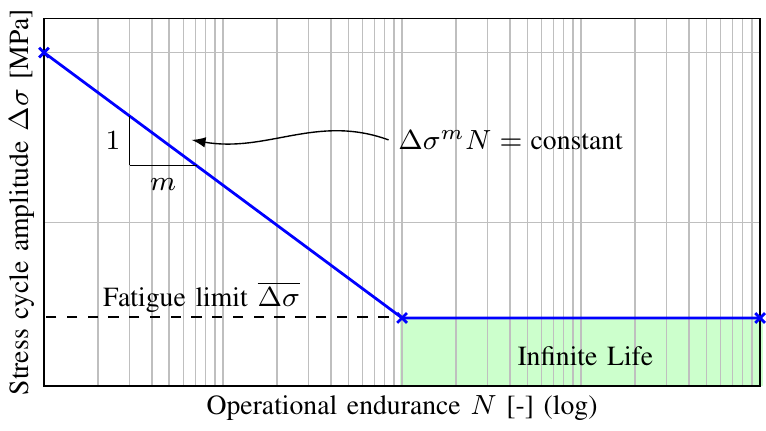}\\
    \caption{Example of SN curve of a ferrous material.}
    \label{fig:illustrativesn}
\end{figure}

The penstock fatigue is estimated using the method reported in \cite{Dreyer2019DigitalCF}, which foresees the following steps:
\begin{enumerate}
    
    \item evaluation of the penstock head, $h_i(t)$, at all penstock's elements, $i=1, \dots, I$;
    
    \item the head is converted into mechanical stress, $\sigma_i(t)$. For open-air penstock, the following model is used (e.g., \cite{Dreyer2019DigitalCF, article}):
    \begin{align}
\sigma_i(t) = (h_i(t) - z_i) \cdot \frac{k D}{2e} && i=1, \dots, I \label{headtostress}
    \end{align}
where $z$ is the elevation,  $D$ and $e$ are the penstock diameter and wall thickness, respectively,  and $k = g \cdot \rho$ converts from head $H$ in meter to pressure $p$ in pascal, where $g$ is the acceleration of gravity and $\rho$ the water density in kg/m$^3$;
    
    \item cycle counting of the mechanical stress with a rainflow algorithm \cite{ASTM, 2009MSSP...23.2712N}. For each penstock element, this algorithm provides J tuples $(\Delta\sigma_{ij}, n_{ij}), j=1,\dots, J$, one for each identified amplitude of stress cycle $\Delta\sigma_{ji}$, where $n_{ij}$ is the number of cycles with stress amplitude $\Delta\sigma_{ji}$;
    
    \item finally, the cumulative damage index, $D_i$, for each penstock element is computed by applying Miner's rule \cite{Dreyer2019DigitalCF}:
    \begin{align}
        D_i= \sum_{j=1}^J \frac{n_{ij}}{N(\Delta\sigma_{ij})}, && \text{for all } i \label{eq:damageindex}
    \end{align}
    where $N(\Delta\sigma_{ij})$ is the maximum number of cycles that the element can perform with a stress cycle of amplitude $\Delta\sigma_{ij}$; this value is given by the so-called SN, or Wohler's curve (discussed in detail below, an example of which is shown in Fig.~\ref{fig:illustrativesn}). The cumulative damage index in \eqref{eq:damageindex} is used to approximate the residual life of a mechanical component. Values near 1 denote that the mechanical element is near to its end of life; zero, vice-versa.
\end{enumerate}

The SN curve reports the number of cycles that a mechanical component can perform at a given stress. It is typically determined empirically and reported in technical standards, such as the BS7910 for welded structures made of steel \cite{HADLEY2018263}, like penstocks. Curves are reported for different quality categories, which refer to specific fatigue design requirements or the presence of flaws in the material. An example of SN curve is shown in Fig.~\ref{fig:illustrativesn}: the initial log-linear trend refers to Basquin equation
\begin{align}
    \Delta \sigma^{m}N = \text{constant}, \label{eq:basquin}
\end{align}
where $m$ is the log-linear slope, which tells that the number of cycles increases for decreasing stress; after this log-linear trend, ferrous materials (such as steel alloys of penstocks) exhibit a \emph{fatigue limit} ($\overline{\Delta\sigma}$ in Fig.~\ref{fig:illustrativesn}), below which the number of cycles increases drastically.

Developing on the fatigue limit of the penstock is the key notion that is leveraged in this paper to reduce fatigue, as formally explained in the next section. It is worth highlighting that the SN curve is an input of our problem, and its identification is beyond the scope of the paper. In an applicative context, the SN curve and the fatigue limit should be suggested by the plant specialist based on the detailed knowledge of the penstock and plant. 

\section{Model Predictive Control for Fatigue Reduction}
As described above, the SN curve of Fig.~\ref{fig:illustrativesn} indicates that a material can undergo an infinite number of cycles when stress variations are below a certain level, called fatigue limit. In other words, stress variations below the fatigue limit do not virtually impact the penstock's service life. This notion is the idea leveraged in this paper to limit penstock fatigue. This section, the main contribution of this paper, describes how closed-form expressions to constraint the stress in the penstock and reduce fatigue are formulated starting from the the guide vane-to-head linear model introduced in the previous section.

\subsection{Stress constraint}
To implement the requirement that stress variations in the penstock should not exceed the fatigue limit, we constraint the stress in a well specif interval. Formally, this reads as:
\begin{align}
   \sigma_\text{nom}  - \frac{\overline{\Delta\sigma}}{2} \le \sigma_i(t)  \le   \sigma_\text{nom} + \frac{\overline{\Delta\sigma}}{2}
   \label{eqn:stress_diseq}
\end{align}
where $\sigma_i(t)$ is the stress on the penstock element $i$ at time $t$, $\sigma_\text{nom}$ is the nominal stress (i.e., the stress during nominal operating conditions of the plant) and $\overline{\Delta\sigma}$ is the fatigue limit. It is easy to verify that, under \eqref{eqn:stress_diseq}, the largest stress amplitude that can occur is:
\begin{align}
    \sigma_\text{nom} + \frac{\overline{\Delta\sigma}}{2} - \left( \sigma_\text{nom} - \frac{\overline{\Delta\sigma}}{2} \right) \le \overline{\Delta\sigma} \; \Rightarrow \; \overline{\Delta\sigma} \le \overline{\Delta\sigma}
\end{align}
thus attaining the requirement of operating below the fatigue limit.

\subsection{Stress constraint as a linear function of the plant's set-point}
Stress inequalities \eqref{eqn:stress_diseq} can be equivalently expressed in terms of penstock head by using the linear relationship in Eq.~\eqref{headtostress}. The reformulated constraints read as:
\begin{align}
    \underline{h} \le h_i(t) \le \overline{h} \label{eq:fconstraints}
\end{align}
where the upper and and lower bounds of the head are 
\begin{align}
    \underline{h} = h_\text{nom} -  \dfrac{\overline{\Delta\sigma} e}{kD} \\
    \overline{h} = h_\text{nom} +\dfrac{\overline{\Delta\sigma} e}{kD} 
   \label{eqn:head_diseq}
\end{align}

By way of the linearized state-space model in \eqref{eq:ss}, the penstock head can be expressed as a linear function of the state and control history. For example, the one-step-ahead prediction at time interval $t$ of the head at penstock's element $i$ is:
\begin{align}
    h_i(t+1) = C_i \left( A \boldsymbol{x}(t) + B_y y(t) + B_z  \boldsymbol{z}(t)\right), \label{eq:onestepahead}
\end{align}
where $C_i \in \mathbb{R}^{1 \times (2I + 1)}$ is an output matrix properly designed to extract $h_i$ from the state vector, and $A, B_y, B_z$ are discrete-time state-space matrices obtained by discretizing \eqref{eq:ss}.

Equations \eqref{eq:fconstraints} and \eqref{eq:onestepahead} are the building blocks used in the next paragraph for the formulation of the MPC optimization problem. As it will be explained next, the need to include predictions in the problem, as in \eqref{eq:onestepahead}, stems from hydraulic dynamics, that determines transients of mechanical loads within the penstock.

\subsection{Formulation of the MPC problem}\label{sec:mpc}
Let $y^\star(t)$ be the guide vane set-point determined by a standard turbine governor (as in Fig.~\ref{fig:hpp_diag}) at time $t$ as a function of the plant regulation duties (primary and secondary frequency control). As $y^\star(t)$ might be unaware of the accumulated effects due to fatigue, we want to find a new guide vane set-point $y^o(t)$ that respects the fatigue limit of the penstock. As mentioned in Section II, the new set-point, $y^o(t)$, should feature the following two attributes:
\begin{enumerate}
    \item it should respect the fatigue limit of the penstock;
    \item it should be as close as possible to $y^\star(t)$ so as to not deviate significantly from the regulation duties of the plant.
\end{enumerate}
These two requirements are formulated in a constrained optimization problem. The first requirement (fatigue limit) is formulated using the linear inequalities discussed in the former paragraph. The second requirement can be expressed in the sense of distance minimization between $y^o(t)$ and $y^\star(t)$.

Being the head within the penstock a dynamic quantity (in the sense that it responds to differential equations), a control decision at a certain time interval influences the head in the future (\footnote{Head dynamics can also be interpreted in the light of the water hammer effect: suddenly closing the guide cause the water to slow down and a shock wave, which travels at a finite speed inside the penstock (thus head variations) and is reflected back and forth between the guide vane and reservoir until dissipated.}). Due to these dynamics, formulating stress constraints for the penstock requires estimating future values of the head, thus motivating a predictive approach. The look-ahead time of the prediction horizon is denoted by $T$; it is chosen as it will be discussed at the end of this section.


Assuming to be at time interval $t$, the set-point is found by solving the following optimization problem:
\begin{subequations}\label{eq:opt}
\begin{align}
\boldsymbol{y}^o = \underset{\boldsymbol{y} \in \mathbb{R}^{T+1}}{\text{arg min}} \left\{ \sum_{\tau=t}^{t+T} \left( y(\tau) - y^\star(\tau) \right)^2 \right\} \label{eq:cost}
\end{align}
subject to guide vane limits
\begin{align}
& 0 \le y(\tau) \le 1, && \tau=t,\mydots,t+T
\end{align}
and penstock model and stress constraints, starting from a known initial condition $x(t)$:
\begin{align}
& h_i(\tau+1) =  C_i \left(A \boldsymbol{x}(\tau) + B_y y(\tau) + + B_z  \boldsymbol{z}(\tau)\right) \label{eq:rec}\\ 
& \underline{h} \le h_i(\tau) \le \overline{h} \label{eq:constr}
\end{align}
\end{subequations}
for the whole optimization horizon $\tau=t,\mydots,t+T$ and all the penstock's elements $i=1,\dots,I$.

In the spirit of MPC, problem \eqref{eq:opt} is applied in a receding horizon fashion, which consists in solving the problem at time $t$ for the whole horizon $t+T$, actuating the first element of the decision vector, and disregarding the rest; at the next time interval, $t+1$, the problem is solved again with updated information and the control is actuated with the same procedure. Also, the model linearization is repeated to account for updated operative conditions of the plant. 

Given that at each time interval a new control action is recomputed, look-ahead time $T$ can be determined by considering that a change of set-point performed at the current time will not impact any longer on the system state after a certain time if the system is asymptotically stable. This consideration provides a formal condition to set $T$ and can be determined analytically.

%

By virtue of the linearized HPP constraints, the optimization problem in \eqref{eq:opt} is convex and can be solved efficiently with off-the-shelf optimization libraries. An analysis of the problem's computational performance is reported in the results section. 

It is worth highlighting that, when solving the optimization problem, the set-points $y^\star(t+1), \dots, y^\star(t+T)$ in \eqref{eq:cost} are not known because they refer to future time intervals. So, they are replaced with persistence predictions, which assumes that future values are the same as the current realization (i.e., $y^\star(t+\tau) = y^\star(t), \tau=1, T$). The use of a persistent predictor is motivated by the fact that the guide vane depends on the grid frequency deviations, which are hard to forecast. Despite its simplicity, it provides satisfactory performance and successful stress reduction, as shown in the Results section.

\section{Methodology for performance evaluation}
\subsection{Case study} \label{sec:cases}
The case study is a 230 MW medium-head HPP with a net-head of 315 meters equipped with one Francis turbine, an open-air 1'100 meter-long penstock, and main characteristics as in Table~\ref{tab:hpp}. The plant is equipped with the standard governor shown in Fig.~\ref{fig:hpp_diag}, which includes a proportional-integral (PI) regulator with a speed droop and set-point for speed-changer setting. 
The governor is a standard PI controller as in \cite{kundur1994power}. 
The PI gains are determined using Ziegler-Nicholas method, which robustness and applicability regardless of the mechanical power of the Francis turbine has been shown in \cite{landry_methodology_nodate}.
Then, the governor is validated in compliance with the ENTSOE qualification tests for PFR as indicated in the standard in \cite{Test}.
The regulator has limits for the rate-of-change and magnitude of the guide vane actuator. The permanent speed droop is set to 2\%. Compared to conventional speed droops for HPPs that are between 2.5\% and 5\%, we choose a lower value to reproduce future operational settings where larger flexibility might be required from dispatchable resources. 

The plant is modeled with the non-linear model discussed in \ref{sec:Modelling}. The penstock is discretized in $I=20$ elements, sufficient to provide an accurate representation of the hydraulic transients. 
The synchronous generator torque is modeled with a second-order model , as in \cite{kundur1994power}, with electrical torque function of the generator's power angle. Rotor dynamics are simulated with the swing equation.
The grid is modeled as an infinite bus, where the grid frequency is imposed by the rest of the power system under the assumption that its size is significantly larger than the simulated plant. 
Grid frequency variations are reproduced considering real system frequency measurements from \cite{EPFL-CONF-203775} of the European interconnected system. 

\begin{table}[!ht]
\renewcommand{\arraystretch}{1.05}
\centering
\caption{Parameters of the HPP case study}\label{tab:hpp}
\begin{tabular}{| l|c|c|c|c|c|}
\hline
\bf Parameter & \bf Unit & \bf Value \\
\hline
Nominal power & MW & 230 \\
Nominal head & m & 315\\
Nominal discharge & m$^3$/s & 85.3\\
Nominal speed & rpm & 375\\
Nominal torque & Nm & 5.86$\times 10^6$\\
Length of penstock & m&  1'100\\
Diameter of penstock & m & 5\\
Wave speed & m/s & 1'100\\
\hline
\end{tabular}
\end{table}

\subsection{Methodology for the numerical simulations}
\begin{figure}[!h]
\centering
\includegraphics[]{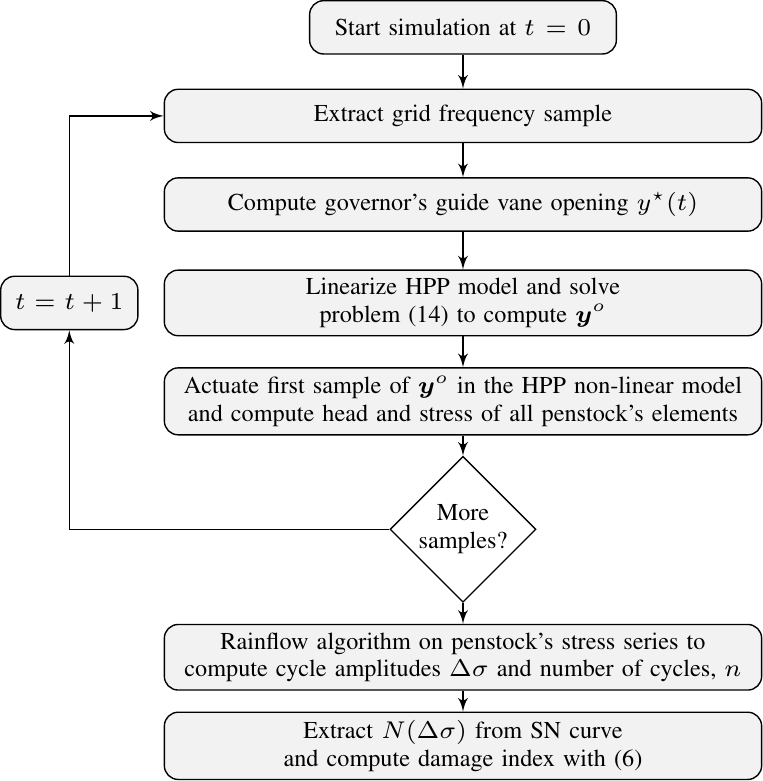}\\
\caption{Procedure for the numerical simulations.}
\label{fig:flowchart}
\end{figure}

The numerical simulation procedure is summarized in Fig.~\ref{fig:flowchart}. At each time interval, the value of the grid frequency is read from a measurement vector and used to compute the guide vane set-point with the HPP governor. Then, a linearized HPP model is computed considering the plant's current working point and used to solve the MPC optimization problem to find the new guide vane set-point. 
The plant non-linear model is then used to compute the ground-truth values of head and stress in the penstock, which are finally used with a given SN curve to assess the damage index. Discretized state-space matrix are computed with 4th order Runge-Kutta. The number of samples $T$ of MPC's look-ahead time problem corresponds to 2~sec.

The fatigue is assessed considering the SN curve with characteristics as in Table~\ref{tab:Q5}, which refers to the quality category Q5 from \cite{HADLEY2018263}. After the fatigue limit, instead of modeling an infinite number of cycles as in Fig.~\ref{fig:illustrativesn}, we model a change of the log-linear slope (from 3 to 5, as reported in Table~\ref{tab:Q5}) to avoid over-optimistic results and, at the same time, consider an empirical uncertainty about on the penstock fatigue limit.

\begin{table}[!ht]
\renewcommand{\arraystretch}{1.05}
\centering
\caption{Parameters of the SN curve}\label{tab:Q5}
\begin{tabular}{|c|c|c|}
\hline
\bf Parameter & \bf Unit & \bf Value \\
\hline
Basquin equation \eqref{eq:basquin}'s slope $m$ &  - &  3 \\
Effective fatigue limit $\overline{\Delta\sigma}$&  MPa &  23 \\
Basquin equation's slope $m$ after $\overline{\Delta\sigma}$ &  - &  5\\
\hline
\end{tabular}
\end{table}

\subsection{Performance metrics}
MPC performance is assessed in terms of incurred penstock damage, measured with the damage index in \eqref{eq:damageindex}. It is convenient to measure performance in terms of improvement with respect to a base case where MPC is not used, namely when the plant operates in its classical configuration (i.e., $y^\star$ in Fig.~\ref{fig:hpp_diag} is the guide vane reference and the controller is bypassed). To this end, we define, for each penstock element $i$, the relative damage index (RDI) as
\begin{align}
    \text{RDI}_i = \dfrac{D_i^{\left(\text{MPC}\right)}}{\max\limits_{i}\left( D_i^{\left(\text{base case}\right)}\right)}, \label{eq:rdi}
\end{align}
where $D_i^{\left(\text{MPC}\right)}$ and $D_i^{\left(\text{base case}\right)}$ are the damage indexes achieved with MPC and in the base case, respectively. In \eqref{eq:rdi}, the reason for dividing by the maximum damage index along the penstock instead of the damage index at $i$ is to avoid that small values of the damage index in certain penstock segments (thus, with negligible impact on the fatigue) generate large value (but insignificant) performance improvements.

A second metric is to compare the performance of different controllers (described in \ref{benchmark}) and concerns evaluating the controllers' ability to track the original guide vane $y^\star$ in the attempt to preserve the original regulation effort. To this end, we use Pearson's correlation coefficient (CC) to measure the similarity between $y^\star$ and $y$ as
\begin{align}
    \text{CC} = \frac{\text{cov}(y, y^\star)}{\sqrt{\text{var}(y)}\sqrt{\text{var}(y^\star)}}. \label{eq:cc}
\end{align}
CC ranges between -1 to +1, where -1 indicates anticorrelation, 0 no correlation, and 1 perfect correlation between the signals.

\subsection{Benchmark controllers}\label{benchmark}
Two controllers from the literature are considered.

\subsubsection{Low-pass filter (LPF)}
A solution commonly advocated in the literature to reduce fatigue is pre-processing the grid frequency signal feeding the governor with a low-pass filter (LPF) to avoid frequent variations of the guide vane set-point, as in \cite{7514942}. The key parameter to be determined is the LPF's cut-off frequency: small cut-off frequencies will significantly smooth the input grid frequency (thus reducing damage), reducing, however, the regulation to the grid; high cut-off frequencies will preserve good regulation capability but without achieving damage reduction. Choosing the cut-off frequency can be done empirically based on, for example, numerical simulations. For this performance comparison, we choose a single-order linear low-pass filter. As discussed in the next section, its cut-off frequency is determined to achieve the same regulation performance as the MPC for a fair comparison.


\subsubsection{Non-linear filter with fatigue limit}
This method, proposed in \cite{9209857}, uses the same notion as in this paper that stress cycles above the fatigue limit produce negligible damage. In order to compute guide vane set-points conducive to low-stress values, it uses a grid frequency-to-penstock stress transfer function model (derived from \cite{kundur1994power}), a filter to trim large stress values, and the inverse filter to reconstruct a new grid frequency signal that respects stress limits.

For this simulation, the value $\overline{\Delta\sigma}$ is chosen as the middle-quality category of \cite{HADLEY2018263}, shown in Table~\ref{tab:Q5}. In a real-life application, this parameter can be adjusted by the plant specialist to reflect the actual fatigue limit of the penstock.



\section{Results}\label{sec:results}

\begin{figure*}
\centering
{\footnotesize
\includegraphics[]{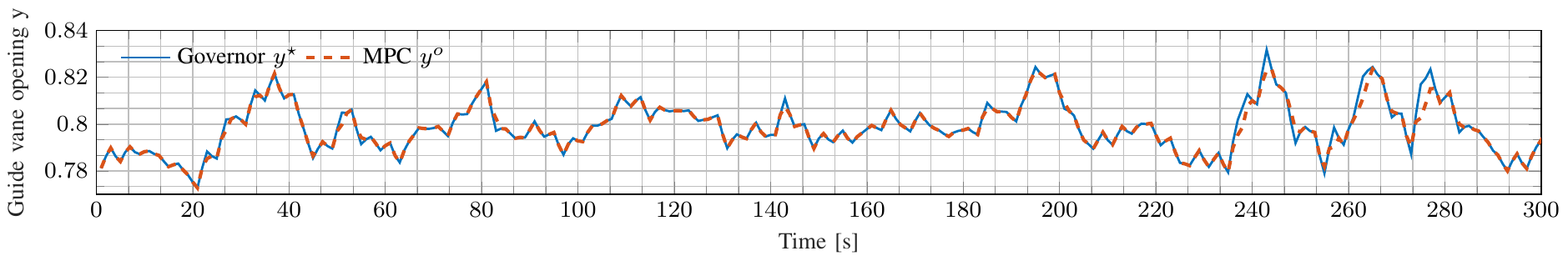}
\includegraphics[]{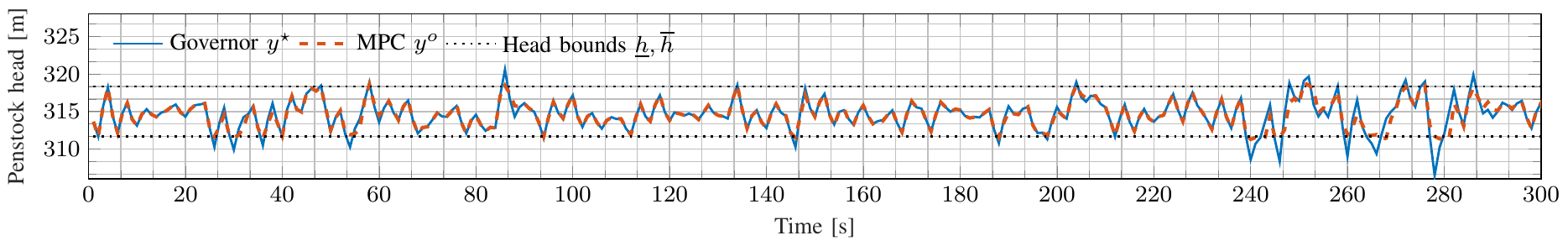}
}
\caption{Set-point actuated by the governor and MPC (top panel) and respective head (bottom panel).}\label{fig:MPC_GV}
\end{figure*}

\begin{figure}
\centering
\footnotesize
\includegraphics[]{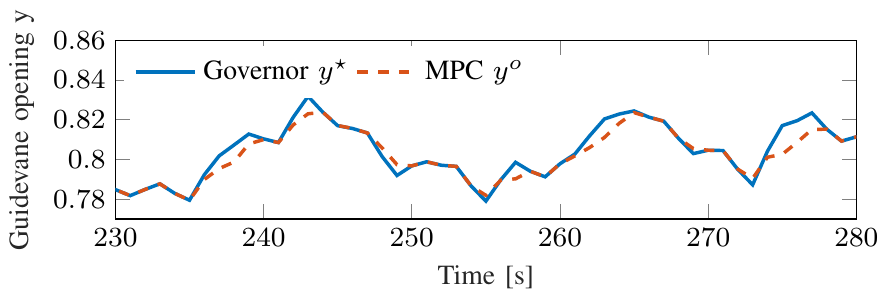}
\caption{MPC actuated guide vane set-point.}\label{fig:MPC_short}
\end{figure}

\begin{figure*}
\centering
\footnotesize
\includegraphics[]{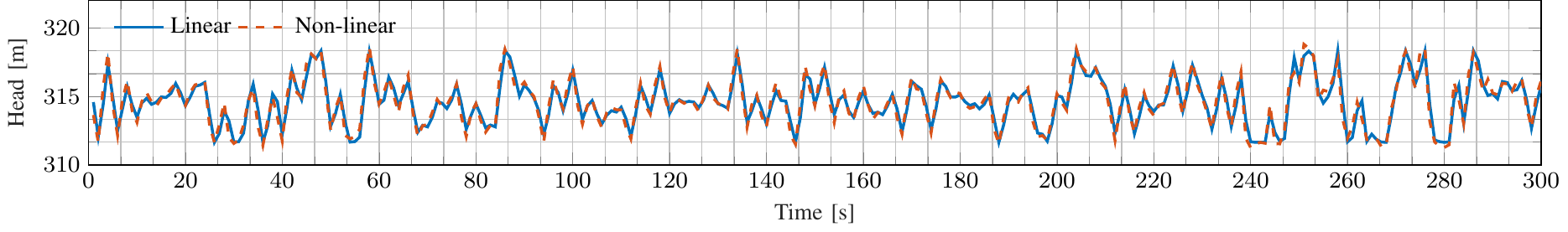}
\caption{Comparison between linear and non-linear head in the critical portion of the penstock.}\label{fig:accuracy}
\end{figure*}

\subsection{Performance assessment of the MPC controller}
The MPC is applied to the output of the HPP governor (as in Fig.~\ref{fig:hpp_diag}) and computes a new guide vane set-point by solving problem \eqref{eq:opt}. This set-point is such that it should respect the penstock's stress constraints. These results are now illustrated.

The top panel of Fig.~\ref{fig:MPC_GV} shows the reference guide vane opening $y^\star$ (blue line) determined by the HPP governor and the one computed by the MPC $y^o$ (dashed red line). These two signals are nearly identical, except for a few cases, as it will be soon described. The bottom panel of Fig.~\ref{fig:MPC_GV} compares the head in the penstock's most critical element resulting from applying $y^\star$ and $y^o$, along with the head limits (dashed lines). 

By looking at the top and bottom panel of Fig.~\ref{fig:MPC_GV}, 
it can be seen that when the original set-point $y^\star$ does not engender violations of the head limits,the MPC set-point $y^o$ is identical to $y^\star$. This follows directly from the formulation of the optimization problem in \eqref{eq:opt}: in particular, when the head constraints in \eqref{eq:constr} are not activated, the problem is unconstrained; its optimal solution happens when $y^\star$ equals $y^o$, resulting in a value of the cost function of 0. However, when constraints becomes active, the optimization problem needs to satisfy the stress constraints and produces a set-point $y^o$ which does not match any longer with $y^\star$.

Fig.~\ref{fig:MPC_short} shows a zoomed view of the guide vanes of the top panel of Fig.~\ref{fig:MPC_GV} for a period when the head limits are exceeded. From this figure, it can be observed that the control action of the MPC resembles a rate limiter.

Fig.~\ref{fig:accuracy} shows a comparison between the linear head estimates and the ground-truth ones simulated with a non-linear model. The relative mean absolute error of the linear estimates is less than 1\% in the range of variations of the guide vane input of a $\pm$0.05 pu.

Finally, Fig.~\ref{fig:lpf_mpc} shows the RDI in \eqref{eq:rdi} along the penstock. Indeed, as discussed in Sec.~\ref{sec:cases}, the penstock is discretized in 20 elements, and the damage is evaluated for each of these elements individually.
It can be seen that the most damaged penstock element is the fifth one, corresponding to a position along the penstock of 200 meters from the upper reservoir. 

As visible, the MPC substantially reduces the relative damage index along all the penstock compared to the case with the standard governor and against a benchmark controller from the literature, a low-pass filter. This last comparison will be discussed more specifically in the next section.

The MPC's convex optimization problem in \eqref{eq:opt} was computed with an average execution time of 22~milliseconds and a standard deviation of the execution time of 5~milliseconds on a laptop with an Intel 5 processor. As the guide vane control action is updated each second, these metrics denote that the problem can be solved with real-time requirements with reasonable margins and thus suitable to be implemented in real-life controllers.


\subsection{Performance comparison against benchmark controllers}
The comparison between controllers consists in evaluating i) the reduction of RDI in \eqref{eq:rdi} and ii) their capacity in tracking the original regulation signal for primary frequency control, measured with metric CC in \eqref{eq:cc}. For the sake of comparison, it is convenient to refer to a situation where the controllers score a similar value for one metric and assess the improvement by looking at the other. Results in these paragraphs are based on 1-hour long simulations. Given that, in power systems, the grid frequency is controlled, we estimate that grid frequency variations within this time interval are well representative of typical dynamics and that results reflect well grid frequency conditions during normal operations.

\begin{figure}
\centering
\footnotesize
\includegraphics[]{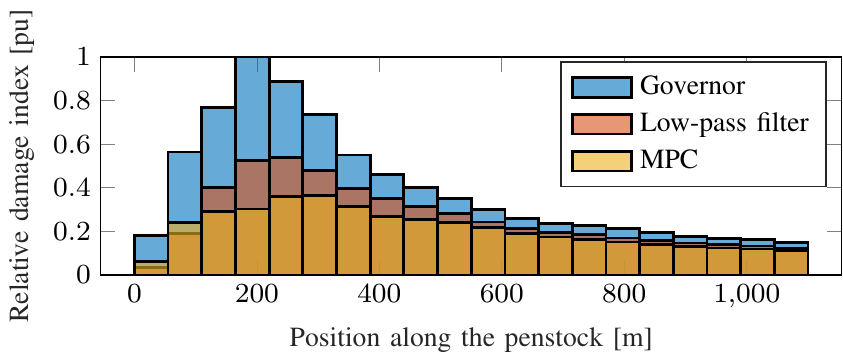}
\caption{RDI along the penstock for original governor, low-pass filter and MPC.}\label{fig:lpf_mpc}
\end{figure}

\paragraph{Low-pass filter versus MPC} We set the cut-off frequency of the LPF to 1.46 Hz to attain the same value of metric CC. In this setting, the LPF achieves to reduce RDI effectively, as shown in Fig.~\ref{fig:lpf_mpc}. However, for the portion of the penstock with the largest damage values (i.e., for penstock segments between 200~m and 400~m), Fig.~\ref{fig:lpf_mpc} shows that the MPC achieves better performance. This result is summarized in the first two entries of Table~\ref{tab:performance}, which show that MPC reduces RDI by nearly 43\% compared to the LPF with the same value of the CC metric.

\begin{figure}
\centering
\footnotesize
\includegraphics[]{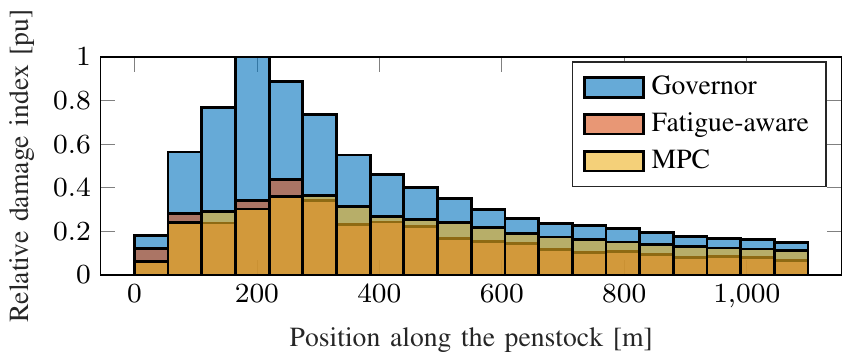}
\caption{RDI along the penstock for original governor, fatigue-aware filter and MPC.} \label{fig:fa_mpc}
\end{figure}

\paragraph{Non-linear filter versus MPC}
The non-linear filter and the MPC are based on the same principle of avoid stress cycles above the fatigue limit, thus they tend to inherently provide similar performance in terms of fatigue reduction. In this context,

In this case, the settings of the non-linear filter are adjusted so as to attain similar reduction of RDI, as shown in Fig.~\ref{fig:fa_mpc}. In these settings, the MPC attain better tracking performance of the original regulation signal, as visible by comparing the correlation coefficients in the last two entries of Table~\ref{tab:performance}. It can be concluded that the MPC achieves, for the same reduction of the damage level, better regulation performance that the non-linear filter. 


\begin{table}[!ht]
\renewcommand{\arraystretch}{1.05}
\centering
\caption{Summary of controller performances}\label{tab:performance}
\begin{tabular}{| l|c|c|c|c|c|}
\hline
\bf Type of controller & \bf CC & \bf RDI (5th penstock element) \\
\hline
Low-pass filter & 0.9948 & 0.53 \\
MPC & 0.9948 & 0.30\\
Fatigue-aware filter & 0.9128 & 0.34 \\
\hline
\end{tabular}
\end{table}

\section{Conclusions}
Primary and secondary frequency regulation duties for HPPs result in increased solicitations on their mechanical components, leading to premature aging and increased maintenance. In high- and medium-head HPPs, frequent and sudden changes of the plant power set-point cause water hammer, damaging the pressurized conduit (penstock) in the long run.

This paper proposed an MPC scheme to model and limit the mechanical stress occurring in the penstock. By leveraging a (linear, to attain a convex formulation of the underlying optimization problem) guide vane-to-head model of the penstock from the literature, the MPC enforces the mechanical load in the penstock to stay below its fatigue limit, which, in damage assessment theory, corresponds to the stress level below which a component can endure a large number of cycles.

To the best of these authors' knowledge, this is the first attempt in the literature to add stress constraints in the control problem of a hydropower plant.

Simulation results developed with a non-linear equivalent circuit model of a medium-head HPP showed that:
\begin{itemize}
    \item linear models provided sufficiently accurate estimates of the penstock head to enforce stress constraints effectively;
    \item the MPC reduces penstock damage compared to standard HPP governor;
    \item the MPC achieves more effective fatigue reduction compared to a benchmark controller based on low-pass filtering the guide vane signal;
    \item the proposed convex formulation of the MPC problem solves fast (tens of milliseconds) and is suitable to be implemented in real-life controllers.
\end{itemize}
The future work is in the direction of experimentally validating the proposed controller and implementing this same notion for fatigue reduction in other kinds of power plants.

\ifCLASSOPTIONcaptionsoff
  \newpage
\fi

\bibliographystyle{IEEEtran}
\bibliography{biblio}

\begin{thebibliography}{10}
\providecommand{\url}[1]{#1}
\csname url@samestyle\endcsname
\providecommand{\newblock}{\relax}
\providecommand{\bibinfo}[2]{#2}
\providecommand{\BIBentrySTDinterwordspacing}{\spaceskip=0pt\relax}
\providecommand{\BIBentryALTinterwordstretchfactor}{4}
\providecommand{\BIBentryALTinterwordspacing}{\spaceskip=\fontdimen2\font plus
\BIBentryALTinterwordstretchfactor\fontdimen3\font minus
  \fontdimen4\font\relax}
\providecommand{\BIBforeignlanguage}[2]{{%
\expandafter\ifx\csname l@#1\endcsname\relax
\typeout{** WARNING: IEEEtran.bst: No hyphenation pattern has been}%
\typeout{** loaded for the language `#1'. Using the pattern for}%
\typeout{** the default language instead.}%
\else
\language=\csname l@#1\endcsname
\fi
#2}}
\providecommand{\BIBdecl}{\relax}
\BIBdecl

\bibitem{iea}
\BIBentryALTinterwordspacing
IEA, ``Hydropower special market report, {IEA},'' Paris, Tech. Rep., 2021.
  [Online]. Available:
  \url{https://www.iea.org/reports/hydropower-special-market-report}
\BIBentrySTDinterwordspacing

\bibitem{yang_wear_2016}
W.~Yang, P.~Norrlund, L.~Saarinen, J.~Yang, W.~Guo, and W.~Zeng,
  ``\BIBforeignlanguage{en}{Wear and tear on hydro power turbines –
  {Influence} from primary frequency control},''
  \emph{\BIBforeignlanguage{en}{Renewable Energy}}, vol.~87, Mar. 2016.

\bibitem{ZHANG2019690}
M.~Zhang, D.~Valentín, C.~Valero, M.~Egusquiza, and E.~Egusquiza, ``Failure
  investigation of a kaplan turbine blade,'' \emph{Engineering Failure
  Analysis}, vol.~97, pp. 690--700, 2019.

\bibitem{LUO2010192}
Y.~Luo, Z.~Wang, J.~Zeng, and J.~Lin, ``Fatigue of piston rod caused by
  unsteady, unbalanced, unsynchronized blade torques in a kaplan turbine,''
  \emph{Engineering Failure Analysis}, vol.~17, no.~1, pp. 192--199, 2010,
  papers presented at the 25th meeting of the Spanish Fracture Group.

\bibitem{surgetank}
S.~Kim, ``Design of surge tank for water supply systems using the impulse
  response method with the ga algorithm,'' \emph{Journal of Mechanical Science
  and Technology}, vol.~24, pp. 629--636, 02 2010.

\bibitem{en12132527}
W.~Wan, B.~Zhang, X.~Chen, and J.~Lian, ``Water hammer control analysis of an
  intelligent surge tank with spring self-adaptive auxiliary control system,''
  \emph{Energies}, vol.~12, no.~13, 2019.

\bibitem{inproceedings}
A.~Adamkowski and M.~Lewandowski, ``Preventing destructive effects of water
  hammer in hydropower plant penstocks,'' 09 2015.

\bibitem{Nicolet2010EvaluationOP}
C.~Nicolet, R.~Berthod, N.~Ruchonnet, and F.~Avellan, ``Evaluation of possible
  penstock fatigue resulting from secondary control for the grid,''
  \emph{Proceedings of HYDRO}, 2010.

\bibitem{Dreyer2019DigitalCF}
M.~Dreyer, C.~Nicolet, A.~Gaspoz, D.~Biner, S.~Rey-Mermet, C.~Saillen, and
  B.~Boulicaut, ``Digital clone for penstock fatigue monitoring,'' in \emph{IOP
  Conference Series: Earth and Environmental Science}, vol. 405, no.~1.\hskip
  1em plus 0.5em minus 0.4em\relax IOP Publishing, 2019.

\bibitem{9160666}
T.~Mäkinen, A.~Leinonen, and M.~Ovaskainen, ``Modelling and benefits of
  combined operation of hydropower unit and battery energy storage system on
  grid primary frequency control,'' in \emph{2020 IEEE EEEIC / I CPS Europe},
  2020.

\bibitem{7514942}
W.~Yang, P.~Norrlund, L.~Saarinen, J.~Yang, W.~Zeng, and U.~Lundin, ``Wear
  reduction for hydropower turbines considering frequency quality of power
  systems: A study on controller filters,'' \emph{IEEE Transactions on Power
  Systems}, vol.~32, no.~2, 2017.

\bibitem{ISGT}
S.~Cassano, F.~Sossan, C.~Landry, and C.~Nicolet, ``Performance assessment of
  linear models of hydropower plants,'' in \emph{2021 IEEE PES Innovative Smart
  Grid Technologies Europe (ISGT Europe)}, 2021.

\bibitem{electric}
O.~Jr, N.~Barbieri, and A.~Santos, ``Study of hydraulic transients in
  hydropower plants through simulation of nonlinear model of penstock and
  hydraulic turbine model,'' \emph{IEEE Transactions on Power Systems},
  vol.~14, 12 1999.

\bibitem{Nicolet:98534}
C.~Nicolet, ``Hydroacoustic modelling and numerical simulation of unsteady
  operation of hydroelectric systems,'' Ph.D. dissertation, EPFL, Lausanne,
  2007.

\bibitem{fatigue}
{L. Cui, I. Frenkel, A. Lisnianski}, \emph{Stochastic Models in Reliability
  Engineering (1st ed.)}.\hskip 1em plus 0.5em minus 0.4em\relax CRC Press,
  2020.

\bibitem{MCBAGONLURI2005291}
F.~McBagonluri and W.~Soboyejo, ``Mechanical properties: Fatigue,'' in
  \emph{Encyclopedia of Condensed Matter Physics}, F.~Bassani, G.~L. Liedl, and
  P.~Wyder, Eds.\hskip 1em plus 0.5em minus 0.4em\relax Oxford: Elsevier, 2005.

\bibitem{adam}
A.~Adamkowski, M.~Lewandowski, and S.~Lewandowski, ``Fatigue life analysis of
  hydropower pipelines using the analytical model of stress concentration in
  welded joints with angular distortions and considering the influence of water
  hammer damping,'' \emph{Thin-Walled Structures}, vol. 159, 12 2020.

\bibitem{article}
A.~Pachoud, R.~Berthod, P.~Manso, and A.~Schleiss, ``Advanced models for stress
  evaluation and safety assessment in steel-lined pressure tunnels,''
  \emph{International Journal on Hydropower and Dams}, 09 2018.

\bibitem{ASTM}
A.~International, ``{ASTM E1049-85(2005), Standard Practices for Cycle Counting
  in Fatigue Analysis},'' Internet Requests for Comments, {ASTM International,
  West Conshohocken, PA,}, {RFC} 1654, 2005.

\bibitem{2009MSSP...23.2712N}
A.~{Nies{\l}ony}, ``{Determination of fragments of multiaxial service loading
  strongly influencing the fatigue of machine components},'' \emph{Mechanical
  Systems and Signal Processing}, vol.~23, no.~8, Nov. 2009.

\bibitem{HADLEY2018263}
I.~Hadley, ``Bs 7910:2013 in brief,'' \emph{International Journal of Pressure
  Vessels and Piping}, vol. 165, 2018.

\bibitem{kundur1994power}
P.~Kundur and N.~Balu, \emph{Power System Stability and Control}, ser. EPRI
  power system engineering series.\hskip 1em plus 0.5em minus 0.4em\relax
  McGraw-Hill, 1994.

\bibitem{landry_methodology_nodate}
C.~Landry, C.~Nicolet, J.~Gomes, and F.~Avellan, ``Methodology to determine the
  parameters of the hydraulic turbine governor for primary control,'' {Power
  Vision Engineering}, Tech. Rep., 2019.

\bibitem{Test}
M.~Scherer, D.~Schlipf, and W.~Sattinger, ``{Test for primary control
  capability},'' Internet Requests for Comments, {Swissgrid}, {RFC}, 2011.

\bibitem{EPFL-CONF-203775}
M.~Pignati, M.~Popovic, S.~Barreto, R.~Cherkaoui, G.~D. Flores, J.-Y.
  Le~Boudec, M.~Mohiuddin, M.~Paolone, P.~Romano, S.~Sarri \emph{et~al.},
  ``Real-time state estimation of the epfl-campus medium-voltage grid by using
  pmus,'' in \emph{2015 IEEE Power \& Energy Society Innovative Smart Grid
  Technologies Conference (ISGT)}.\hskip 1em plus 0.5em minus 0.4em\relax IEEE,
  2015.

\bibitem{9209857}
S.~Cassano, C.~Nicolet, and F.~Sossan, ``Reduction of penstock fatigue in a
  medium-head hydropower plant providing primary frequency control,'' in
  \emph{2020 55th UPEC}, 2020.

\end{thebibliography}

\end{document}